\documentstyle{article}
\advance\textheight by1in
\advance\textwidth by1in
\advance\voffset by-0.5in
\advance\hoffset by-0.5in
\def\sec#1\par{\let\Large=\relax\section{#1}}
\def\subsec#1\par{{\let\large=\relax\let\bf=\it\subsection{#1}}}
\def\bi{\begin{list}{$\bullet$}{\topsep0pt \parsep=\parskip \itemsep=0pt}}
\def\ei{\end{list}}
\def\[{$$}
\def\]{$$}
\def\mede#1{\left\langle#1\right\rangle}

\begin{document}
\noindent HLRZ preprint 46/93

\noindent Univ.\ Cologne Appl.\ Math.\ WP 93.133
\baselineskip16pt \tolerance2000
\title{Deterministic models for traffic jams}
\author{
Kai~Nagel${}^{+*}$, Hans~J.~Herrmann${}^*$
\\~\\
\normalsize ${}^+$ ZPR and Math.\ Inst.\ der Universit\"at, Weyertal~86-90,
\\ \normalsize 50923~K\"oln 41, Germany (permanent address)
\\ \normalsize {\tt kai@mi.uni-koeln.de}
\\~\\
\normalsize ${}^*$ HLRZ, c/o KFA, Pf.~1913, 52425~J\"ulich, Germany
}
{{\def\newpage{\relax}
\maketitle
}}
\noindent {\bf Abstract}

We study several deterministic one-dimensional traffic models.
For integer positions and velocities we find the typical high and
low density phases separated by a simple transition. If positions and
velocities are continuous variables the model shows self-organized
criticality driven by the slowest car.
\sec Introduction

Everybody knows of the seemingly erratic motions of cars jammed on
highways. One wonders whether they are due to a random behaviour
of the individual drivers or if there is an intrinsic chaotic
mechanism. In favour of the first hypothesis is the existence of
regular kinematic waves in dissipative systems with excluded
volume~\cite{Leib93}. For this reason many traffic models include
rather important statistical noise in time~\cite{NaS92}. In favour of
the second hypothesis are measurements performed on Japanese
highways showing a $1/f$ spectrum in the Fourier transformed
density fluctuations~\cite{Mush9?} which might stem from some
self-organized criticality~\cite{BTW88}. It is therefore interesting to
see if a traffic model without noise is able to give the
observed erratic behaviour and its $1/f$ spectrum.

In this paper, we systematically investigate deterministic variants of
a simple one-dimensional model for traffic jams, which is, on the one
hand, astonishingly realistic~\cite{NaS92} and which can, on the other
hand, be used for unprecedented high speed microscopic traffic
simulations on parallel and/or vector computers~\cite{NaS93}.  Further
analysis of the nondeterministic model may be found
in~\cite{ScS93,INSS93}, and simplifications of this modeling technique
have been applied to two-lane traffic~\cite{Nag93} as well as to
two-dimensional gridlock problems~\cite{BML92,CMMS93,Taka}.

The deterministic model investigated in this paper shows, under certain
circumstances, trivial critical behavior at a non-trivial critical
density, and, providing enough input and output of mass at the
boundaries, the open version of the model selects this density
automatically.  In addition, a straightforward extension of this model
to floating point numbers leads to a deterministic system which
exhibits non-trivial complex behavior even when starting from a totally
ordered initial state.

\sec The prototype model

The prototype of all models presented in this paper is a
one-dimensional horizontal road with traffic flowing to the right
of length~$L$ (representing a road) with
periodic boundary conditions.  Each box may either be empty, or it may
be occupied by a particle (vehicle) with an integer speed 0, 1, 2,
$\ldots$.  A particle with velocity $v$ is either accelerated to $v
\to v+1$ when there are at least $v+1$ empty boxes in front of the
particle; or it is slowed down to $v \to k$ where $k$ is the number of
empty boxes in front of the particle:
\[
\vtop{\tabskip2em
\halign{$#$ & $#$ & $#$ \cr
v \le \Delta x -2 & \leadsto & v \to \Delta x -1 \cr
v \ge \Delta x    & \leadsto & v \to v + 1  \ ,\cr
}}
\]
  where $\Delta x = k + 1$ is the gap between the particles plus the
size of one particle.   After the velocity update, the particle is
advanced $v$ boxes to the right (``propagation step'').

The original model as described in~\cite{NaS92} goes beyond this model
insofar as it introduces a maximum velocity (cf.~further down in the
present paper) and as it adds a high degree of noise to the dynamics
taking into account the fluctuations of everyday driving behavior.

For the model presented here, three types of deterministic updates are
possible: parallel, right circular, and left circular.  Parallel means
that all $N$ vehicles in the system are treated simultaneously.  In
this case, a vehicle cannot advance to more than one box behind the
car in front has been the time step before.  This update rule leads to
gaps between particles proportional to their distance, which is
realistic for traffic dynamics~\cite{May90}, and it therefore
tends to disperse the particles (Fig.~1) as if they had a
repulsive force of a range proportional to their speed.

``Right circular'' update means that the update initially starts at
some arbitrary particle (usually the leftmost one), adjusts its speed
and performs its motion, and then proceeds with the next particle to
the right and so on.  When having periodic boundary conditions, this amounts to
an ``update demon'' which circulates through the system.
Nevertheless, for the ``circular right'' update it turns out that it
is equivalent to the parallel update, with the following exception: At
the position of the ``update demon'', it is possible to have two
particles close together at high speed; in this case the left one just
has been moved and the right one is the next particle to be moved.
But this behavior does not change the mean properties for large system
sizes.

A ``left circular'' update, however, produces a markedly different
dynamics.  Here, a particle moves when the car in front already has moved
away, which means that particles may move in clusters, each particle
just catching up with the car in front during its update.  In practice, {\em
any\/} initial configuration leads
after some time to only one cluster containing all
vehicles, moving at speed $v_{end} = L-N$
(cf.~Fig.~2).  It is this update which we will denote as
``circular'' in the following.

\sec Transient times and damage times for the prototype model

We measured each model's average ``transient time'' $\tau_t$, which is
the time the system needs to settle down from a random initial
condition to a regular steady state.  Random initial condition here
means that we placed the particles randomly and then drew integer
velocities randomly between $0$ and $L-1$ for each particle.  ($L-1$
is an upper bound for the velocity: Even in the case of only one
particle in the system, when all update methods are equivalent, due to
the periodic boundary conditions the particle ``sees'' itself at
$\Delta x = L$ and can therefore not go faster than with $v = \Delta x
-1 = L - 1$.)

With parallel update, the steady state is reached when the entire
configuration of the system at time $t+1$ is a simple left shift of the
configuration at time $t$ (as in the lower part of
Fig.~1), which means that each particle assumes the
velocity of the car in front.
Nevertheless, already this simple model contains some of the
features of a high traffic regime: If one follows the trajectory of one
vehicle individually, regions of relatively free movement are
alternating with regions of high density and slow speed. For the
parallel update rule the average speed in the steady state equals the
number of empty sites divided by the number of particles $\mede{v} =
(L-N) / N$.

For the left circular update however, the steady state consists, as
mentioned above, of only one cluster of particles which all move with the
highest possible speed $L-N$.

In addition, we measured a ``damage time'' $\tau_d$ in order to
estimate the duration of a disturbance\cite{13}.  To do so, after reaching the
steady state, we reduced the velocity of one randomly chosen particle
by the smallest possible amount: $\tau_d$ is then the time the system
needs to reach again a new steady state.  Technically, for the parallel
update, we reduced the velocity of the randomly chosen particle by one
just before the propagation step, whereas we reduced it by two {\em
after\/} the propagation step for the left circular update.  In the latter
case,
a reduction by only one would have had no consequences at all.

Numerical results for $\tau_t$ as a function of the density $\rho = N
/ L$ are given in Fig.~3.  We find for the parallel update and $L \to
\infty$ the scaling $\tau_t \propto \rho^{-1}$ and $\tau_d \propto
\rho^{-1}$, i.e.\ diverging times
for the density going towards a ``critical'' density of zero.

The left circular update data (Fig.~4) shows no scaling
behavior in $\rho$; and in fact there are simple explanations for the
data:  Since  $\tau_t$ is simply the time the initially slowest particle
needs to reach its final velocity, one has
  \[
\tau_t \propto \mede{v_{max}} - \mede{\min(v_{ini})}
= (L-N) - {L \over (N+1)} = {L - \rho \, L - 1 \over 1 + 1/\rho L}
\sim L \ .
\]
The damage time $\tau_d$ is unity, when the leading particle is
disturbed, and three otherwise.  In average, this leads to $\tau_d = 3
- 1/N$.

\sec Introduction of a speed limit

For real highway traffic as well as for dissipative systems under a constant
force it is usually true that they have a limiting velocity even for
the ``free'' particles which do not feel any influence of neighbors.
Introducing a maximum velocity in our models simply means that a
particle is only accelerated when it has not yet reached the maximum
speed.  We will use a maximum velocity of $5$ throughout this article.  It
should be noted that only this introduction of the maximum velocity
makes the particle interaction short-ranged.

When again starting from random initial conditions, this leads with
parallel update for high densities to configurations similar to the
case without velocity limit, but for low densities a new phase appears
having clusters of vehicles moving with maximum velocity
(Fig.~5).  In other words, there are gaps between
particles moving with maximum speed; a gap technically being defined
as one or more sites not being crossed or touched by a particle during
the propagation step. A left circular update, however,  gives intermittent
clusters of maximum velocity for all densities.  As there is no
velocity-dependent repulsive force between particles, many of the
particles stick together.

The transient times for the parallel update case are plotted in
Fig.~6.  We find that at the non-trivial density $\rho_c =
1/6$ the transient time $\tau_t$ shows a remarkable peak which grows
with system size; finite size scaling analysis confirms that it diverges as
  \[
\tau_t(\rho_c) \propto L \ .
\]
The same is true for the damage time: $\tau_d(\rho_c) \propto L$.  For
the left circular update, the main change with respect to infinite velocity
is that now even the transient time $\tau_t$ does no longer depend on
$L$ for $L \to \infty$.

The phenomenological reason for the divergence at $\rho_c = 1/6$ for
the parallel update is as follows (explained for the damage time): At
low densities, a disturbance only travels upstream to the next gap;
the relevant length scale obviously depends on $\rho$ and not on $L$
(see Fig.~5).  For high densities, however, (compare Fig.~1), the
disturbance introduces a new wave front or reinforces an older one,
and it leaves a gap traveling downstream because the disturbed
particle cannot catch up quickly enough the car in front.  This gap
can only be filled in a region of particles which neither accelerate
nor move at maximum speed, i.e., in a region with higher density.  The
length scale therefore depends on $\rho$.  In between the high and the
low density regimes, a transition takes place; and the critical point
is a state where particles move with maximum velocity $v_{max}=5$ and
minimum distance $\Delta x_{min} = v_{max} + 1 = 6$ (and therefore
$\rho_c = 1/\Delta x_{min} = 1/6$).  If one introduces a disturbance
into this ``critical'' state, a fronts moves contrary to the flow
direction until it meets downstream (periodic boundary conditions!)
the gap caused by the disturbance before; hence the $L$-dependance of
$\tau_d$ at $\rho_c$.

At the same time, $\rho_c$ is the density which corresponds to maximum
particle throughput (current)~$j = \rho \, v$.  In fact, in the low
density regime all particles have maximum speed $v_{max}$ and therefore
$j_< = v_{max} \, \rho$, whereas in the high density regime the
average speed is equal to the number of empty sites divided by the
number of particles and therefore
  \[
j_> = {L-N \over N} \, {N \over L} = 1 - \rho \ .
\]
  These two straight lines intersect at $\rho_c = 1 / (1 + v_{max})$,
which is therefore the density corresponding to maximum throughput.

For convenience, the different regimes are summarized in the following
table:
\newdimen\tmpsize  \tmpsize0.3\hsize \tabskip1em plus2em
$$\vtop{\halign{#\hfill&&\vtop{\hsize\tmpsize\tolerance8000
\parindent0pt\baselineskip12pt#}\cr
		& parallel update 	& left circular update	\cr
\noalign{\medskip\hrule\medskip}
$v_{max} = L$
& Tendency for equidistance, waves. (*)
& All cars in one cluster.
\cr
\noalign{\medskip}
$v_{max} = 5~,~\rho > \rho_c$
& Tendency for equidistance; waves with $v < v_{max}$. Similar to (*).
& Tendency to cluster, but not all cars together; all cars $v_{max}$. (**)
\cr
\noalign{\medskip}
$v_{max}=5~,~\rho < \rho_c$
& Formation of gaps; no waves; all cars $v_{max}$.
& Tendency to cluster, but not all cars; all cars $v_{max}$. Similar to (**).
\cr
}}$$

\sec Self-organization of the critical state

If one assumes a very dense jam as in Fig.~7, then the
outflow from this jam assumes exactly the critical configuration of
particles moving at maximum speed 5 and with a distance $\Delta x$
of~6.  This is no longer true for an open boundary (Fig.~8); but
it can be restored, e.g.\ by forcing a higher acceleration for the
first particle (Fig.~9).  In this last case, a disturbance
(as in Fig.~9) travels with constant speed to the left
boundary, whereas the gaps travels with constant speed to the right
boundary.  Conceptually, this case of self-organized criticality
therefore belongs to BTW's one-dimensional sandpile model~\cite{BTW88},
with the difference that our model selects a state of non-trivial
density.

With respect to reality, this means that the outflow of the traffic
jam adjusts itself exactly at the critical density and therefore at maximum
capacity, and due to the criticality, new disturbances
have long-ranged effects through density waves traveling with
constant speed. This may no longer be true for a
bottleneck situation (e.g.\ a two-lane directional road merging into
only one lane): The open boundary described in Fig.~8
leads to lower density and throughput, and the same is true for a
more realistic model with strong noise~(\cite{NaS92,Nag93}).  Which of
these findings exists in reality with its mixture of different
vehicles and drivers is still an open question, although everyday
experience tells us that traffic {\em inside\/} a bottleneck is indeed
relatively stable and should therefore be below the critical density.

\sec A continuous version

In order to investigate if our model shows more complex (and therefore
more realistic) behavior when introducing a higher number of degrees
of freedom,
we considered a continuous version of our model.  The one-dimensional
system still has length~$L$ with periodic boundary conditions; but
velocity $v_i$ and position $x_i$ of a vehicle $i$ are now continuous
variables. The update rule is as follows:\bi

\item
If the velocity is high with respect to the gap, then the car slows down:
\[
v > \Delta x - \alpha \quad
\leadsto \quad v \to \max( \, 0 \, , \, \Delta x - 1 \, ) \ ;
\]
(the ``max'' is only necessary to prevent negative velocities);

\item
else if the velocity is low with respect to the gap {\em and\/} slower
than five, then the car accelerates:
\[
v < \Delta x - \beta \ \ \&\ \ v < v_{max} \quad
\leadsto \quad v \to v + \min( \, 1 \, , \, \gamma*\Delta x \, ) \ .
\]
Note that this rule allows maximum speeds up to nearly six.

\item
After the velocity has been updated for all vehicles according to the
last two rules, we move all vehicles simultaneously according to
their velocities.

\ei
In our simulation
we used $\alpha = 0.5$, $\beta = 3.0$, and $\gamma = 0.1$.

The only new feature of this model with respect to the integer version is
that the acceleration is weaker when the distance is still small.  A
distinct feature of this model which it shares with the integer model
is that there is a ``dead zone'' in the distance to the next car ahead
where a driver neither accelerates nor slows down.  This is in
accordance with psychological investigations insofar as it stresses
the importance of physiological {\em thresholds\/} in order to make
human drivers react.  Many car-following models use the velocity
difference to the vehicle ahead instead of the gap as primary
stimulus~\cite{May90}, and a practical implementation~\cite{Wie74}
uses thresholds which depend on the velocity difference {\em and\/}
thresholds which account for the gap between two cars.  And indeed,
the latter are necessary in order to prevent grossly unrealistic
behavior, allowing, e.g., very small gaps at very high velocities.
However, suppression of the former, as we have done for our model,
still leads to essentially realistic behavior.

With this model, we performed simulations of different setups.
Whereas the normal closed systems showed a behavior similar to the
integer model (i.e., settling down to an ``imitate your
leader''-state), already the introduction of one slightly slower
vehicle leads to complex and unpredictable behavior.  To be specific,
the setup is as follows.  In a system of length~$L$ and with periodic
boundary conditions, we initially placed $N = [ \rho \cdot L
]$~vehicles on sites $1$ to $N$, all with velocity zero, where
$\rho$ was chosen small enough to prevent any effect of the last car
of the platoon on the first one through the periodic boundary
conditions.  (A platoon is an ensemble of vehicles travelling
together.)  Starting from this totally ordered initial state, the
system was allowed to evolve according to the above rules, with
the exception of the first vehicle, the speed of which
was kept fixed at $v_{lead}=4.99999$ after its initial
acceleration. This is a
simplification of the well known situation where a number of fast
vehicles has to follow a slower one which they cannot pass, and
besides its obvious single-lane applications this situation also
occurs on freeways when many passenger cars want to pass a
group of trucks.

Fig.~10 shows a section of the evolution of the system.  For a certain
time, the density behind the first car gets larger because all vehicles
close up.  But at seemingly random instances, many of the followers
have to slow down in one large collective event, which redistributes
the vehicles with a lower density.  As the first vehicle moves freely,
the simulation represents, in spite of the periodic boundary
conditions, the situation of a platoon moving infinitely in space.

In Fig.11 we see the time evolution of a system which had been
transformed to the coordinates of the first vehicle, i.e. the
positions of all cars are given relative to the first car. We see that
equally spaced cars rapidly evolve into a fluctuating state (right
hand side). In this new state density increases give rise to very
short bursts (traffic jams) of very different sizes which redistribute
the cars backwards such that in some cases they even start again in
equally spaced patterns.

Although the model itself is totally deterministic, small
perturbations may lead to totally different trajectories in phase
space due to the chaotic dynamics.  We have noticed this effect when
simulating the same number of vehicles in systems of slightly
different size: After a certain time, the development of the systems
diverges, due to small differences in the transfer of cars through the
periodic boundary.  In order to clarify that this divergence is an
intrinsic consequence of the dynamics and not just the enhancement of
numerical round-off errors, we have compared single precision with
double precision calculations.  The overall result of
these tests is that noise enters the system with the same ``speed''
for all cases, which is a strong indication that complex behavior
originates from a chaotic dynamics and is not driven by the limited
precision of the floating point numbers.

In addition, the principal behavior of the model (i.e.\ the formation
of the collective shocks) is robust with respect to parameter
changes. More precisely, we could not find a
qualitatively different behavior for
changes of the parameters $\alpha$, $\beta$, $\gamma$,
and $v_{lead}$ within the following range
$0.1 \le \alpha \le 0.6$, $2.0 \le \beta \le 5.0$,
$0.08 \le \gamma \le 0.12$, $4.5 \le v_{lead} \le 4.99999$.

In order to quantify these observations, we measured the distribution
of times $\tau$ between consecutive ``braking''
events for the last vehicle ($\tau$ is the time from the end
of one braking to the beginning of the next).  Braking
is defined here as a slowing down according to the rules for the
velocity update.  We performed simulations on a Parsytec GCel-3,
replicating a system with a fixed number of vehicles but different
system sizes on up to 512~processors and averaging the results.
For instance for $N=1900$ vehicles, we waited about $3
\cdot 10^5$~time steps to let the transients die out, and then measured
the distribution of $\tau$ during about $1.1 \cdot 10^6$ further time
steps.  This specific simulation took about $33$ hours on
256~processors.

According to Fig.~12, this distribution displays a remarkable
straight line on a log-log-plot, fulfilling the power law
\[
n(\tau) \propto \tau^{-\alpha}
\]
with $\alpha = - 2.2 \pm 0.1$. This non-trivial exponent is a strong
indication for the existence of self-organized criticality\cite{BTW88}
for this model.

Many aspects of this model are remiscent of the so-called train model
for earthquake dynamics~\cite{Maria}.  Instead of pulling at one end,
the slower car may be seen as {\em pushing\/} against the other cars
which want to move faster.  This leads to a slowly increasing average
density, and at some time this density locally exceeds a critical
threshold.  The reaction is a more or less drastic slowing down of the
corresponding vehicle, which may or may not force the next vehicle to
slow down as well.  By this mechanism, avalanches of all sizes are
generated, which may propagate through the entire platoon of vehicles.

\sec Conclusion

We have studied several one-dimensional deterministic traffic models.
As long as the velocities and positions are integers we find the
well-known \cite{NaS92,CMMS93}, rather simple situation of a low
density phase with particles having maximum velocity, a high density
phase of low velocity waves and a simple transition between
the two. When the velocity limit diverges the low density phase
vanishes.

Much richer is a model in which continuous positions and velocities
are allowed. If one car is a little slower we observe the spontaneous
creation of erratic traffic jams of all sizes and a power law
spectrum in the times between them. This self-organized behaviour
is reminiscent of Burridge and Knopoff's original ``train
model''\cite{Maria}. It would be interesting to analyze this model
within the general theoretical considerations concerning
 self-organized criticality\cite{Kru91,CCGS90}.

\sec Acknowledgements

We acknowledge computer time on the Parsytec GCel-3 of the ZPR
(University of Cologne). Enlightening discussions with J.~Kert\'esz and
M.~Schreckenberg have helped us to beware of numerical round-off errors.
KN thanks for discussions with and
motivation by A.~Bachem, J.~Lee and M.~Leibig.

\parindent0pt \parskip0.3\baselineskip
\vfill\eject\noindent{\bf Figure captions}

{\bf Fig.~1:}
Parallel update, $\rho = 0.27$, $v_{max} = L = 70$. The entire highway
is shown. Consecutive lines show configurations at consecutive time
steps (i.e., time is evolving downwards), after the velocity update
and before the propagation step.  Empty spaces are marked by dots,
particles by their integer velocities (``{\tt *}'' means that velocity
is larger than 9). Particles move from left to right.  The figure
shows the evolution of the system from a random initial configuration
to five steps after reaching the steady state and then the effect of a
disturbance.  Except for the first line, the same evolution could be
possible for a system with $v_{max} = 5$.

{\bf Fig.~2:}
Left circular update, $\rho = 0.3$, $v_{max} = L = 70$. The entire system is
shown.  The figure shows the evolution of the system starting from a
random initial configuration.

{\bf Fig.~3:} Transient times~$\tau_t$ for $v_{max} = \infty$ using
parallel update.  Systems of the same size are connected by lines.

{\bf Fig.~4:}
  As Fig.~3, for left circular update.

{\bf Fig.~5:}
  Parallel update, $v_{max}=5$, $\rho = 0.1$ (below $\rho_c$),  $L=70$.
This regime has no counterpart in the
dynamics without speed limit.

{\bf Fig.~6:}
  Transient times $\tau_t$ for $v_{max} = 5$ (parallel update).
Systems of the same size are connected by lines.

{\bf Fig.~7:}
Deterministic outflow from a traffic jam.

{\bf Fig.~8:}
Acceleration from a dense situation (open system).  Particles are inserted with
maximum
speed at the left boundary whenever there is space, but due to the other
particles ahead they are immediately slowed down.  In consequence, the
system is not able to reach its maximum throughput.

{\bf Fig.~9:}
Acceleration from a dense situation (open system) with a forced acceleration of
the
leftmost particle.  This means that at this position, a particle
instantly can accelerate from zero to, e.g., the maximum speed five
(see, e.g., from the first to the second line).

{\bf Fig.~10:} Evolution of the continuous model
(5600~time steps), $N = 61$, $L = 1024$.  As
in the pictures for the deterministic system, consecutive lines
represent configurations at consecutive time steps; vehicles are
denoted by black squares, empty spaces are left blank.

{\bf Fig.~11:} First 4200~time-steps of the continuous model,
transformed to the coordinates of the leading (rightmost) vehicle.
Only every third time-step is plotted, everything else is as in
Fig.~10.  The jam-waves are now clearly visible as nearly horizontal
lines. The flow is from left to right.

{\bf Fig.~12:} Distribution $n(\tau)$ of times $\tau$
between consecutive braking events
for $N=190$ and $N=1900$ cars.  (The times are
collected in ``logarithmic bins'', so that the $y$-axis is
proportional to $\tau \, n(\tau)$.)

\end{document}